# Visualization approach to assess the robustness of neural networks for medical image classification

Elina Thibeau-Sutre[a], Olivier Colliot[a], Didier Dormont[a,b], Ninon Burgos[a]

[a] Institut du Cerveau et de la Moelle épinière, ICM, Inserm U 1127, CNRS UMR 7225, Sorbonne Université, Inria, Aramis project-team, F-75013, Paris, France ; [b] AP-HP, Hôpital de la Pitié Salpêtrière, Department of Neuroradiology, F-75013, Paris, France

## 1. INTRODUCTION

Deep learning methods have shown a high performance potential for medical image analysis [1], particularly classification for computer-aided diagnosis. However, explaining their decisions is not trivial and could be helpful to achieve better results and know how far they can be trusted. Many methods have been developed in order to explain the decisions of classifiers [2]–[7], but their outputs are not always meaningful and remain difficult to interpret. In this paper, we adapted the method of [8] to 3D medical images to find on which basis a network classifies quantitative data. Indeed, quantitative data can be obtained from different medical imaging modalities, for example binding potential maps obtained with positron emission tomography (PET) or gray matter (GM) probability maps extracted from structural magnetic resonance imaging (MRI). Our application focuses on the detection of Alzheimer's disease (AD), a neurodegenerative syndrome that induces GM atrophy. We used as inputs GM probability maps, a proxy for atrophy, extracted from T1-weighted (T1w) MRI. The process includes two distinct parts: first a convolutional neural network (CNN) is trained to classify AD from control subjects, then the weights of the network are fixed and a mask is trained to prevent the network from classifying correctly all the subjects it has correctly classified after training. The goals of this work are to assess whether the visualization method initially developed for natural images is suitable for 3D medical images and to exploit it to better understand the decisions taken by classification networks. This work is an original work and has not been submitted elsewhere.

## 2. MATERIALS AND METHODS

### 2.1 Data description and preprocessing

Data used in the preparation of this article were obtained from two public datasets: the Alzheimer's Disease Neuroimaging Initiative (ADNI) database (adni.loni.usc.edu) and the Australian Imaging, Biomarkers and Lifestyle (AIBL) study. We used the T1w MRI available in each of these studies. Two diagnosis groups were considered:

- CN: sessions of subjects who were cognitively normal (CN) at baseline and stayed stable during the follow-up;
- AD: sessions of subjects who were diagnosed as AD at baseline and stayed stable during the follow-up.

The population of ADNI and AIBL are described in Table 1.

Table 1. Summary of ADNI and AIBL participant demographics, mini-mental state examination (MMSE) and global clinical dementia rating (CDR) scores at baseline.

| Dataset | Label | Subjects | Sessions | Age | Gender | MMSE | CDR |
|---|---|---|---|---|---|---|---|
| ADNI | CN | 330 | 1 830 | 74.4±5.8 [59.8, 89.6] | 160 M / 170 F | 29.1±1.1 [24, 30] | 0: 330 |
| ADNI | AD | 336 | 1 106 | 75.0±7.8 [55.1, 90.9] | 185 M / 151 F | 23.2±2.1 [18, 27] | 0.5: 160; 1: 175; 2: 1 |
| AIBL | CN | 429 | 730 | 72.5±6.2 [60, 92] | 183 M / 246 F | 28.8±1.2 [25, 30] | 0: 406, 0.5: 22, 1: 1 |
| AIBL | AD | 76 | 108 | 73.9±8.0 [55, 93] | 33 M / 43 F | 20.6±5.5 [6, 29] | 0.5: 31; 1: 36; 2: 7, 3: 2 |

*Values are presented as mean ± SD [range]. M: male, F: female*

Preprocessing of T1w MR images was performed with the Clinica software platform (www.clinica.run) [9]. First the datasets were converted to the BIDS format, then the t1-volume preprocessing pipeline of Clinica was applied [10]. This pipeline performs bias field correction, non-linear registration and tissue segmentation using the Unified Segmentation approach [11] available in SPM12. The GM maps in MNI space were retrieved for the image analysis.

## 2.2 CNN classification

The following sections describe the evaluation procedure, the hyperparameters selection and implementation details that are linked to the classification of AD vs CN subjects with CNNs. During training, the weights and biases of the network $w$ are optimized to maximize the score function $f_w$ on a set of images $X$ as follows: $w^* = \operatorname*{argmax}_{w} f_w(X)$.

### 2.2.1 Evaluation procedure

The ADNI dataset was split into training/validation and test sets. The ADNI test set consisted of 100 randomly chosen age- and sex-matched subjects for each diagnostic class (i.e. 100 CN subjects, 100 AD patients). The rest of the ADNI dataset was used as training/validation set. We ensured that age and sex distributions between training/validation and test sets were not significantly different. The model selection procedure, including model architecture selection and training hyperparameter fine-tuning, was performed using only the training/validation dataset. For that purpose, a 5-fold cross-validation (CV) was performed, which resulted in one fold (20%) of the data for validation and the rest for training. Note that the 5-fold data split was performed only once for all the experiments with a fixed seed number (*random_state* = 2), thus guaranteeing that all the experiments used exactly the same subjects during CV. The AIBL dataset was used as an independent test set to assess the CNN generalization ability. Test and validation sets included only one session per subject.

### 2.2.2 Hyperparameter selection

We performed a random search [12] to select the architecture and optimization hyperparameters of our CNN. The hyperparameters explored for the architecture were the number of convolutional blocks, of filters in the first layer and of convolutional layers in a block, the dimension reduction strategy (by using a max pooling layer or by setting the stride of the last convolutional layer of the convolutional block to 2), the number of fully-connected layers and the dropout rate. Other hyperparameters such as the learning rate, the weight decay, the batch size, the data preprocessing and the intensity normalization were also part of the search.

Only one experiment was performed per architecture tested using the first split of the CV due to the computational cost of the random search. The chosen architecture was the one that obtained the best balanced accuracy on the validation set. This architecture (displayed in Figure 1) is composed of 7 convolutional blocks followed by a dropout layer and a fully-connected layer. Each convolutional block (C1, C2 or C3) is made of 1 to 3 sub-blocks and a max pooling layer with a kernel size and a stride of 2. Each sub-block is composed of a convolutional layer with kernel size of 3, a batch-normalization layer and a leaky ReLU activation. The predicted label of the input image is the class with the highest output probability.

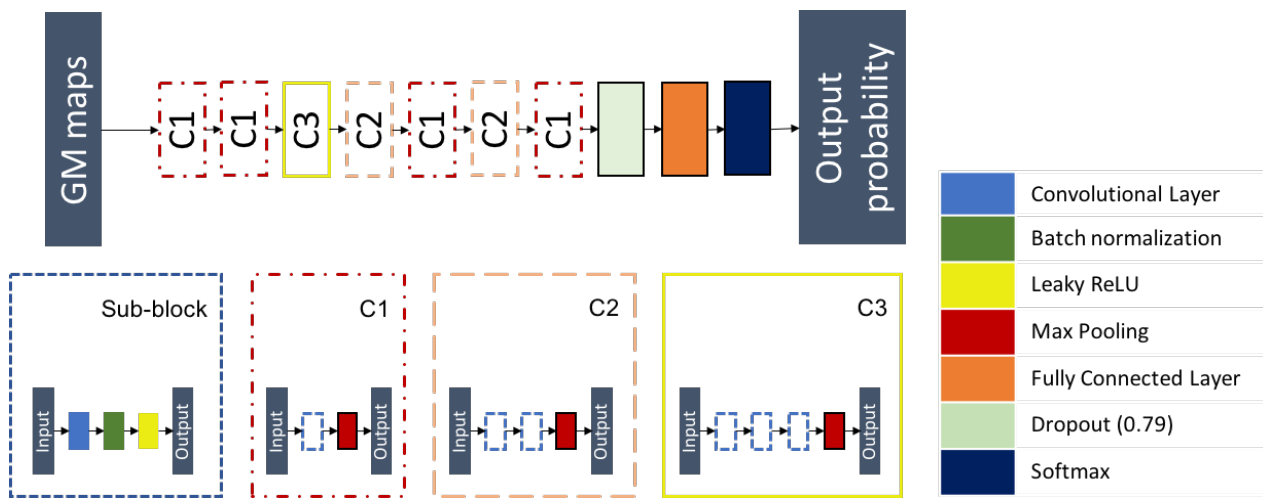

Figure 1: Architecture of the CNN classifier determined following to a random search procedure.

### 2.2.3 CNN training

The weights of the convolutional and fully connected layers were initialized as described in [13], which corresponds to the default initialization method in PyTorch. We applied the following early stopping strategy for all the classification experiments: the training procedure does not stop until the validation loss is continuously higher than the lowest validation loss for $N$ epochs ($N$=5); otherwise, the training continues to the end of a pre-defined number of epochs (30). The training and validation loss were computed with the cross-entropy loss. For each experiment, the final model was the one that obtained the highest validation balanced accuracy during training. The balanced accuracy of the model was evaluated at the end of each epoch.

### 2.3 Visualization method

The proposed visualization method extends the framework of [8]. Once the classification network has been trained, its parameters are fixed to the best value found, denoted as $w^*$. The method consists in computing a mask that will overlay the most meaningful parts of an image to prevent the network from classifying it correctly. In the following, the goal is to mask AD images that were correctly classified by the CNN so that it systematically classifies them with the CN label. The mask $m$ is a 3D volume of the same size as the input image and hide parts of the image in a voxel-wise manner. In this application, each voxel $u$ of the input image $X$ will be masked by a constant value $\mu$ according to the value of the mask for this voxel. The mask values are included in [0, 1]. The masked input image $X'_m$ at voxel $u$ is defined as:

$$X'_m(u) = m(u)X(u) + (1 - m(u))\mu \quad (1)$$

As AD patients suffer from gray matter atrophy, the goal of the masking method would be to artificially simulate gray matter restoration in a minimal number of brain regions to make them look like CN subjects. By setting $\mu = 1$, the mask was trained to artificially increase the probability of GM for the minimum set of voxels which will lead to the maximum decrease of the performance of the CNN. The optimal mask $m^*$ is the mask for which the following loss function is minimized:

$$m^* = \underset{m}{\operatorname{argmin}} \lambda_1 \|1 - m\|_{\beta_1}^{\beta_1} + \lambda_2 \sum_u \|\nabla m(u)\|_{\beta_2}^{\beta_2} + f_{w^*}(X'_m) \quad (2)$$

The first regularization term ensures that a minimum set of voxels is selected while the second ensures that the mask is smooth enough and is not made of scattered voxels.

Once the mask training is finished, values above 0.95 are set to 1. This ensures that the CNN is only perturbed by the zones identified by the mask, and not by the small gradients that can be found on all the surface of the mask.

### 2.3.1 Quality check procedure

As the visualization method is very sensitive to outliers when applied to a group of images, a quality check procedure was performed before mask optimization. Figure 2 displays images that passed or failed this procedure. This quality check includes two steps:

1. The GM maps were sorted in increasing order by their maximal value. Images with a maximal value lower than 0.95 were automatically rejected. 8 sessions were removed during this procedure.
2. One image was removed after training a group mask. During the training of the first group mask this session led to a significant increase of the loss. This image was removed as it suffered from defects (the eyes were segmented as gray matter).

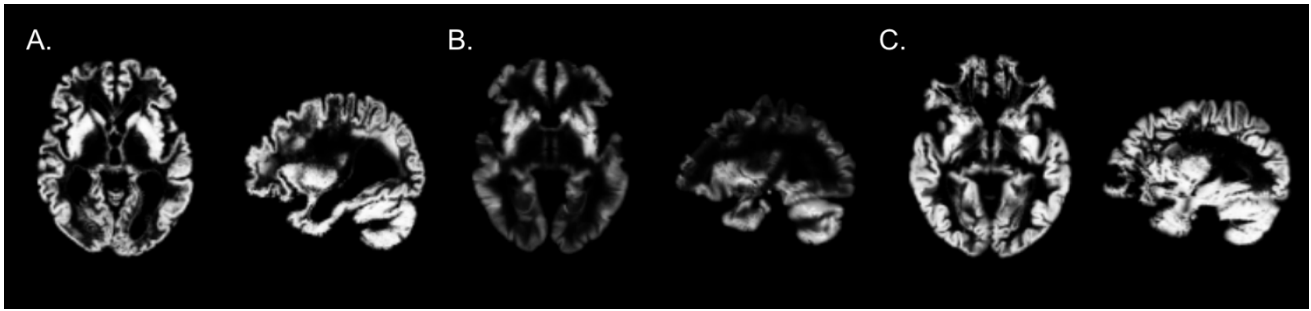

Figure 2: A. Example of image that passed the quality check. B. Example of image removed during the first step of the quality check. C. Image removed during the second step of the quality check.

### 2.3.2 Grid search on visualization hyperparameters

A grid search was performed to choose the set of hyperparameters linked to the computation of the mask: the coefficients for the regularization $\lambda_1, \lambda_2, \beta_1, \beta_2$ in equation (2). The learning rate was arbitrarily fixed to 0.1. The grid search was only performed on the group level masking for AD label.

### 2.3.3 Group level masking

To find the regions most related to Alzheimer's disease according to the CNN, we computed a mask based on the subset of images of AD subjects that were in the training set and validation set and all correctly classified by the network. For AIBL, the subset of all AD sessions correctly classified by the network was used. We exploit the fact that a voxel-wise correspondence exists between the GM maps, thanks to the non-linear registration, to iteratively build a group mask: the mask is initialized with all its values set to 1 and is then updated each time with a different image. The subset of well-classified AD of the validation set used for the CNN training was again used as a validation set for mask optimization. At the end of each epoch, the masking loss was evaluated on this set to save the best mask according to the validation masking loss. To assess the robustness of the CNN training, the masks obtained for different folds (i.e. different input images and initializations) and different runs of the same folds (i.e. same input images but different initializations) were compared by pairs. The mean value of all pairs gave the similarity between folds or runs of the same fold.

### 2.3.4 Session level masking

Masks were also produced at the session level based on a single image. To avoid the overfitting risk due to the use of only one image instead of a set of images as in the previous section, the regularization terms $\lambda_1$ and $\lambda_2$ were multiplied by 100. No validation set was used for these experiments as the goal is precisely to overfit the individual pattern of one image instead of finding a general pattern that may correspond to a group of images. Session level experiments include a longitudinal and cross-sectional analysis. In the longitudinal analysis, all the sessions of one subject were compared by pairs and the mean value of these comparisons gave the intra-subject similarity for this subject. The mean intra-subject similarity is then the mean value of the intra-subject similarity of all AD subjects. For the cross-sectional analysis, the mean value of all pairwise comparisons of baseline sessions of all AD subjects gave the inter-subject similarity measure. These analyses are performed to assess the stability of the visualization method and provide a baseline value by using the inter-subject similarity for the different metrics.

### 2.3.5 Visualization method training

The mask was initialized with a matrix of the same size than the input images (121x145x121) full of ones. We applied a similar early stopping strategy than for the classification experiments: the training procedure for group level masking on ADNI does not stop until the relative difference between the validation loss and the lowest validation loss is superior to a tolerance of 0.05 for a patience of $N$ epochs ($N$=5); otherwise, the training continues to the end of a pre-defined maximum number of epochs (150). For the group level masking on AIBL, the patience was increased to 25 and the maximum number of epochs to 300 as the number of AD subjects is smaller than for ADNI. For the session level masking, the patience was increased to 200 and the maximum number of epochs to 5,000, while the tolerance was decreased to 0.01. The loss corresponds to the argmin argument of Equation (2). For each experiment, the final mask was the one that obtained the lowest validation loss during training. The loss of the mask was evaluated at the end of each epoch.

## 2.4 Metrics of evaluation

The similarity between masks was evaluated in two ways. The output probabilities of the CNN for the true class ($prob_{CNN}$) for an input masked by two masks optimized in two different contexts (e.g. different runs for the group level masking, different sessions of the same subject for the session level masking) are used to establish a comparison based on the CNN perception of the input. A mean output probability close to 1 means that the first model is not perturbed by the mask optimized for the second model, meaning that the two models are dissimilar. A ROI-based similarity was also computed to assess the similarity of two masks according to the 120 regions-of-interest (ROIs) of the AAL2 atlas [14]. For each ROI, 1 minus the sum of the values in the ROI is computed, resulting in a ROI-vector of size 120 for each mask. Each value in the ROI-vector represents the density of the mask in the associated ROI. The ROI-based similarity between two masks is then the cosine similarity of two ROI-vectors. A value close to 1 means that the densities of the masks are the same between the ROIs, a value close to 0 means that the locations of the masks have no intersection.

## 3. RESULTS

Once the architecture was chosen and the CNN was trained on all folds, the classification performance was evaluated on the independent test set to ensure the absence of overfitting. The validation balanced accuracies on the five folds were 0.95, 0.82, 0.96, 0.85 and 0.87, giving an average of 0.89. The test balanced accuracies on the five folds were 0.89, 0.87, 0.90, 0.86 and 0.87, giving an average of 0.88. Moreover, the balanced accuracies obtained on the independent test set AIBL were 0.85, 0.92, 0.91, 0.92 and 0.92, giving an average value of 0.90. We could thus conclude that the network was not overfitting and we could use it for the visualization task.

### 3.1 Grid search on visualization hyperparameters

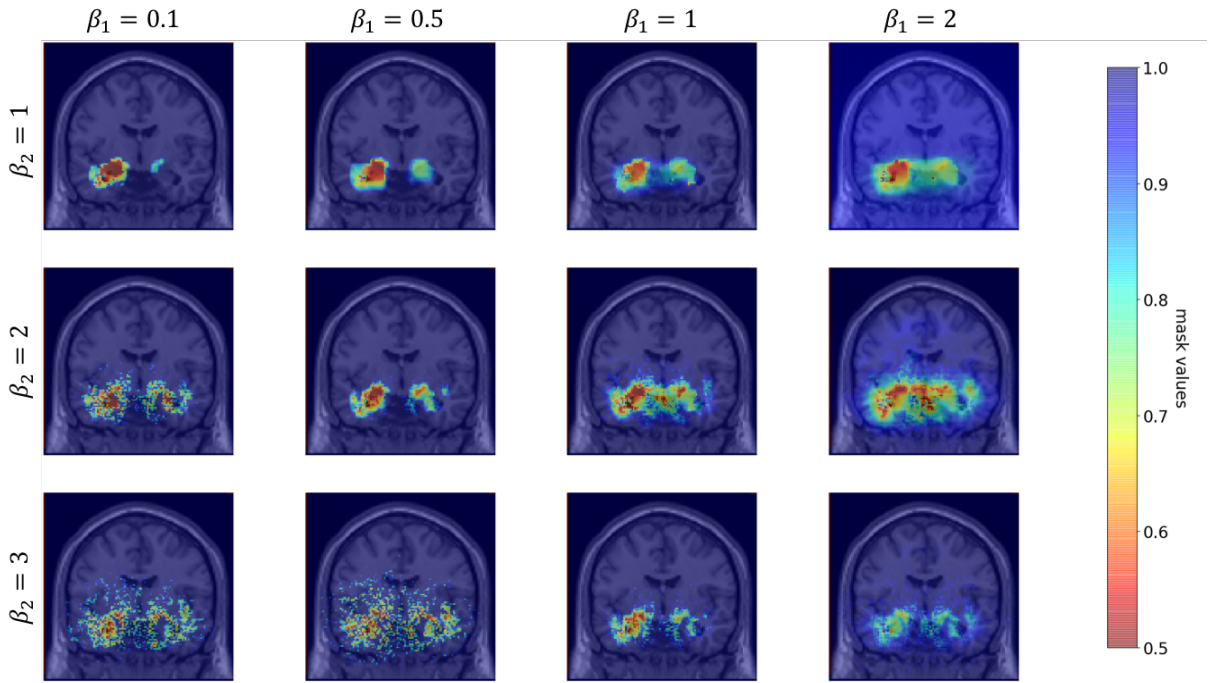

Figure 3: Comparison of masks obtained for different values of the visualization hyperparameters $\beta_1$ and $\beta_2$.

First the hyperparameters $\beta_1$ and $\beta_2$ were chosen with fixed $\lambda_1 = 0.0001$ and $\lambda_2 = 0.001$. The choice of the values $\beta_1 = 0.1$ and $\beta_2 = 1$ was made based on visual inspection. We observe on Figure 3 that when $\beta_1$ decreases, the minimal value of the mask decreases and this prevents from producing a mask with a large set of values close but different from 1. When $\beta_2$ increases, the value of the second term becomes negligible before the first term of equation (2). This leads to a very scattered mask as it is dominated by the first term of the regularization.

The hyperparameters $\lambda_1$ and $\lambda_2$ were then chosen with fixed $\beta_1 = 0.1$ and $\beta_2 = 1$. The choice of the values $\lambda_1 = 0.0001$ and $\lambda_2 = 0.01$ was made based on visual inspection and the stability of the loss during mask training. We observe on Figure 4 that when $\lambda_1$ increases, the surface covered by the mask decreases until it only becomes scattered points. When $\lambda_2$ increases, the surface covered by the mask increases.

### 3.2 Robustness of the visualization method

Different experiments were conducted to assess whether the method was robust enough to help interpret the results of the CNN. Indeed [15] highlighted that some visualization methods developed to interpret the results of neural networks (for example guided back-propagation and guided grad-CAM) did not depend on model parameters, as they gave the same result with a pretrained network or a randomized one. Hence we need to check if the visualization method gives coherent results based on the CNN training.

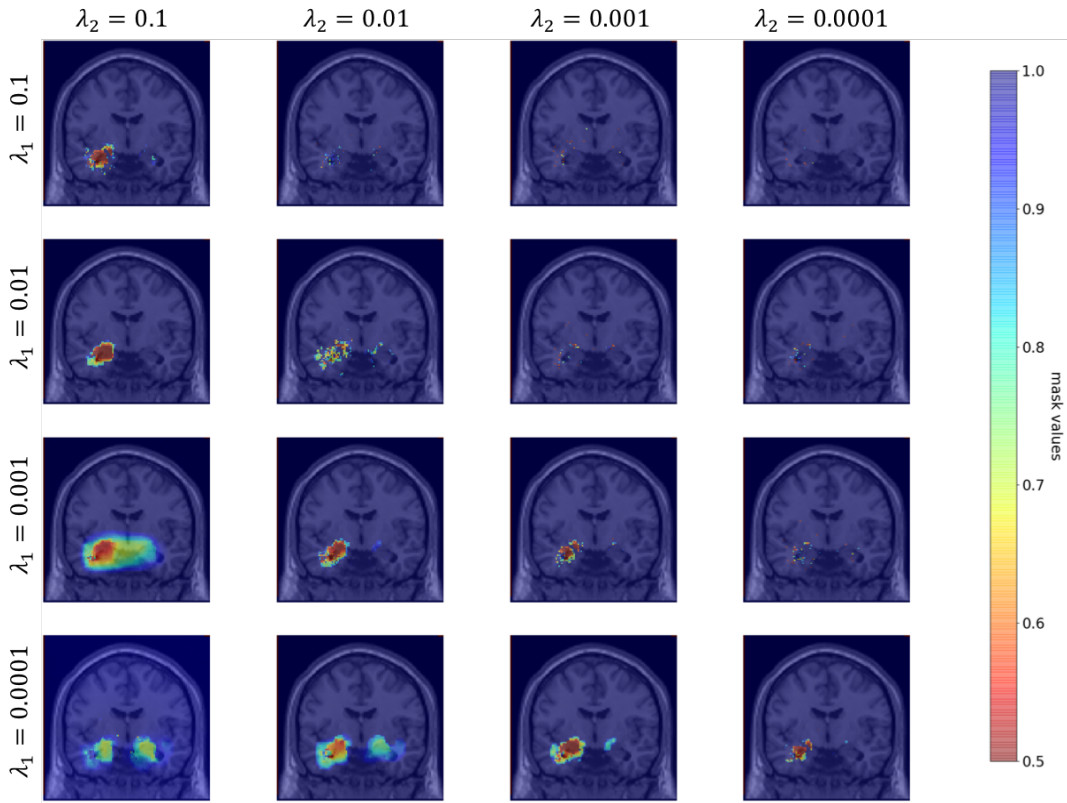

Figure 4: Comparison of masks obtained for different values of the visualization hyperparameters $\lambda_1$ and $\lambda_2$. Note that for $\lambda_1 = 0.0001$ and $\lambda_2 = 0.1$, the learning rate was fixed to 0.01 as the mask optimization did not converge with a learning rate of 0.1.

#### 3.2.1 Group level masking

This experiment aims to assess the coherence of the proposed visualization approach with the a priori knowledge of the disease. One mask was optimized for each of the five models trained on the five folds of the cross-validation. Though these masks do not always overlap, they focus on a set of ROIs known to be particularly affected during AD progression. To confirm this visual observation, the list of the 5 ROIs in which the mask has the lowest values was extracted for each fold. All masks include in this list at least one hippocampus and parahippocampal gyrus. Moreover, the fusiform gyri (4 masks out of 5) and the amygdalae (3 masks out of 5) are frequently highlighted by the masks. Other regions such as the putamen, the pallidum, the inferior temporal gyrus and the thalamus appear only once in these lists.

Moreover, to assess the robustness of the method towards data used for mask optimization we compared the masks obtained by applying the visualization method on ADNI or AIBL data using the five networks trained on the five folds of the CV on ADNI training/validation set. The corresponding masks are displayed on Figure 5. The ROI-based similarities between the pairs of masks were 0.92, 0.99, 0.93, 0.89 and 0.97. These are comparable to the intra-subject ROI-based similarity (0.94). The $prob_{CNN}$ dissimilarities were very small as all the dissimilarities were smaller than $10^{-3}$. This indicates that for a given pretrained network, a mask optimized for images of ADNI (resp. AIBL) correctly occludes the images of AIBL (resp. ADNI). However, the comparison of the masks in this way may not be completely fair. Even though the number of epochs and the patience of the early stopping procedure were increased for AIBL masking, the masks on ADNI and AIBL did not benefit from the same number of iterations. This factor leads to masks that comprise a different number of points, as the effect of the regularization terms is correlated to the number of iterations. This means that though the masks highlight the same locations in the brain, the difference in regularization makes the masks more dissimilar that they would be if we could find an equivalence for the hyperparameters that control the number of epochs (patience, tolerance and maximum number of epochs). Hence the dissimilarity here may not be due to the difference of datasets, but to the different number of iterations done during the mask optimization.

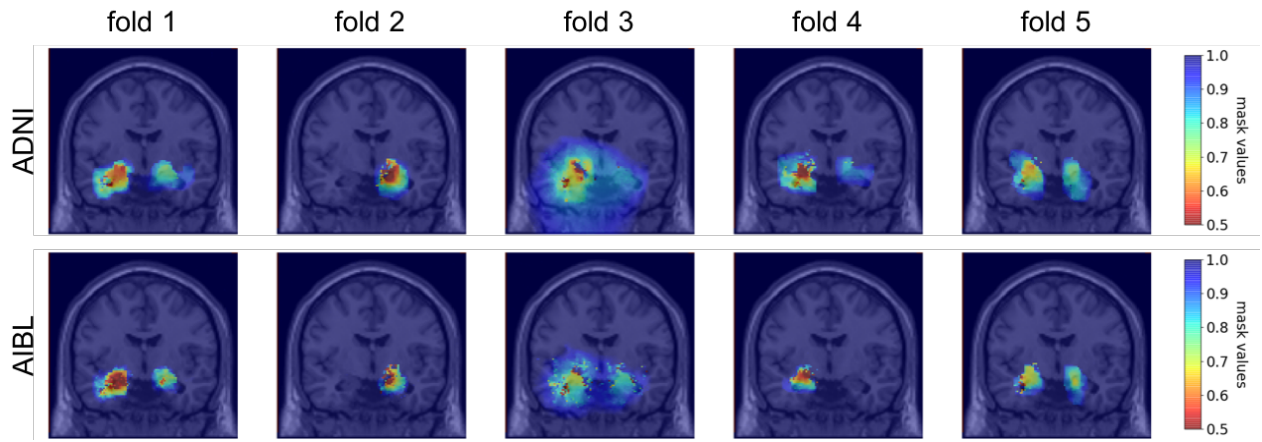

Figure 5: Coronal view of the group masks trained on ADNI (first line) and AIBL (second line). Each column corresponds to a model trained on one fold of the CV on training/validation ADNI set.

### 3.2.2 Session level masking

The inter-subject similarity and dissimilarity were evaluated to 0.80 and 0.58 for the ROI-based and the $prob_{CNN}$ metrics respectively. The intra-subject similarity and dissimilarity were evaluated to 0.94 and 0.11 for the ROI-based and the $prob_{CNN}$ metrics respectively. The higher intra-subject similarity compared to the inter-subject similarity ensures that the visualization metric is robust as the same pattern is generated for different sessions of the same subject.

### 3.2.3 Similarity across hyperparameters

With the ROI-based similarity, we can assess whether the masks produced by varying one hyperparameter value are similar. To observe this similarity, we reused the same masks as those produced for the grid search (see Figure 3 and 4).

First, the similarities using different $\beta_1$ and $\beta_2$ values were computed with fixed $\lambda_1 = 0.0001$ and $\lambda_2 = 0.001$ and a learning rate of 0.1. The similarities between masks produced with $\beta_1$ values in {0.1, 0.5, 1, 2} and fixed $\beta_2 = 1$ are given in Table 2.A. As expected when looking at the masks obtained in Figure 3, the masks are highly similar except for the value $\beta_1 = 2$ for which the first regulation term became negligible in front of the second regulation term in equation (2). It resulted in a very smooth mask which is dense in all regions of the brain as 97.5% of values are below 0.95. This explains why the ROI-based similarity is so low between this mask and the others, though the regions identified seem similar at visual inspection. Other masks have a high similarity (> 0.95 in all cases). The similarities between masks produced with $\beta_2$ values in {1, 2, 3} and fixed $\beta_2 = 0.1$ are given in Table 2.B. There is more variability for this hyperparameter, though the similarity between two consecutive values is still high (> 0.90).

Table 2 A. Similarity across different $\beta_1$ with fixed $\beta_2 = 1$. B. Similarity across different $\beta_2$ with fixed $\beta_1 = 0.1$.

| A | $\beta_1 = 0.1$ | $\beta_1 = 0.5$ | $\beta_1 = 1$ | $\beta_1 = 2$ |
|---|---|---|---|---|
| $\beta_1 = 0.1$ |  | 0.98 | 0.97 | 0.32 |
| $\beta_1 = 0.2$ | 0.98 |  | 1.00 | 0.36 |
| $\beta_1 = 1$ | 0.97 | 1.00 |  | 0.37 |
| $\beta_1 = 2$ | 0.32 | 0.36 | 0.37 |  |

| B | $\beta_2 = 1$ | $\beta_2 = 2$ | $\beta_2 = 3$ |
|---|---|---|---|
| $\beta_2 = 1$ |  | 0.90 | 0.70 |
| $\beta_2 = 2$ | 0.90 |  | 0.91 |
| $\beta_2 = 3$ | 0.70 | 0.91 |  |

The similarities using different $\lambda_1$ and $\lambda_2$ values were then computed with fixed $\beta_1 = 0.1$ and $\beta_2 = 1$ and a learning rate of 0.1. For both hyperparameters the similarity is high between two consecutive values (>0.90), as can be seen in Table 3.

Table 3 A. Similarity across different $\lambda_1$ with fixed $\lambda_2 = 0.01$. B. Similarity across different $\lambda_2$ with fixed $\lambda_1 = 0.0001$

| A | $\lambda_1 = 0.1$ | $\lambda_1 = 0.01$ | $\lambda_1 = 0.001$ | $\lambda_1 = 0.0001$ |
|---|---|---|---|---|
| $\lambda_1 = 0.1$ |  | 0.93 | 0.84 | 0.83 |
| $\lambda_1 = 0.01$ | 0.93 |  | 0.95 | 0.91 |
| $\lambda_1 = 0.001$ | 0.84 | 0.95 |  | 0.91 |
| $\lambda_1 = 0.0001$ | 0.83 | 0.91 | 0.91 |  |

| B | $\lambda_2 = 0.1$ | $\lambda_2 = 0.01$ | $\lambda_2 = 0.001$ | $\lambda_2 = 0.0001$ |
|---|---|---|---|---|
| $\lambda_2 = 0.1$ |  | 0.98 | 0.85 | 0.72 |
| $\lambda_2 = 0.01$ | 0.98 |  | 0.92 | 0.82 |
| $\lambda_2 = 0.001$ | 0.85 | 0.92 |  | 0.96 |
| $\lambda_2 = 0.0001$ | 0.72 | 0.82 | 0.91 |  |

These results highlight the stability of the method toward the hyperparameters choice, as two consecutive hyperparameter values led to two masks with a ROI-based similarity superior to the inter-subject similarity (0.80). Moreover, all the masks involved in this section analysis correctly occlude the CNN perception: for all masks, the mean output probability of the AD class on the validation dataset is below $10^{-6}$.

### 3.3 Robustness of the CNN training

After having assessed the robustness of the visualization method, we applied it to better understand the factors influencing the training process of the CNN classifier based on several scenarios: for different folds (different initialization, different training/validation split) and different runs (different initialization, same training/validation split). Figure 6 displays the masks obtained for the five folds and five runs of the first fold.

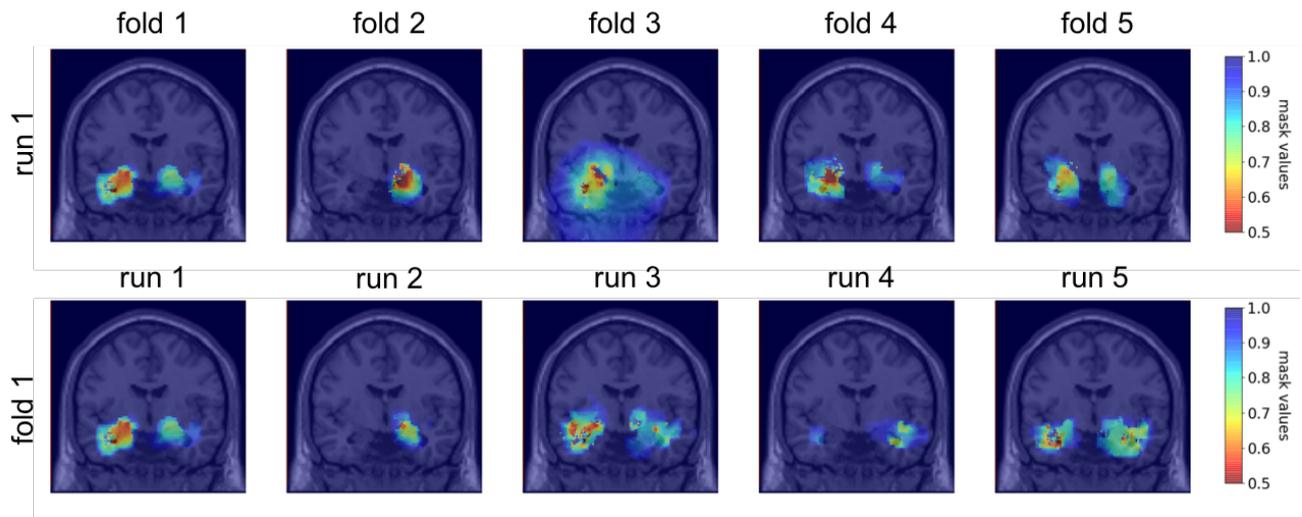

Figure 6: Coronal view of the group masks obtained for the five folds of the CV on the first run (first line) and of the group masks obtained for five runs of the first fold (second line)

With the prob$_{CNN}$ metric using the validation set, the dissimilarity between folds and 5 runs of the same fold are equivalent with respectively 0.78 and 0.82. The similarities computed with the ROI-based metric are also equivalent with respectively 0.65 and 0.69 between folds and between runs of the same fold. This indicates that the impact of the distribution of the data between training and validation is minimal compared to the initialization of the CNN and the training process. Moreover, we observe that the dissimilarity between folds / runs of the same fold is higher that the inter-subject dissimilarity obtained with session level masking. This could mean that the regions on which the CNN relies on to identify the diagnosis mainly depend on the initialization and the training process and that the CNN training is not robust towards the regions identified.

## 4. DISCUSSION

We extended a visualization method to 3D medical imaging data and used it to better understand the decisions made by a classification network.

We first assessed the robustness of the proposed visualization approach. We showed that it gave coherent results as the regions identified by the mask are representative of AD [16]. This coherence is also confirmed by the fact that the intra-subject similarity is higher than the inter-subject similarity. Moreover, the high similarity across neighbouring values of the hyperparameters of the masking method indicates that this method is stable towards hyperparameter selection. Finally, we assessed that the method appears robust towards the data used for the construction of group masks by comparing masks computed using ADNI and AIBL datasets.

We then applied the visualization approach to assess the robustness of the CNN training. We demonstrated that even if the classification performances on the test set are very similar between the different folds, the training of the CNN for our application is not robust as the inter-subject similarity for one training is higher than the similarity between two re-trainings or two folds of the network. This problem of robustness in CNN training may exists for many medical applications in which the number of samples is not sufficient for the network to learn stable meaningful features. This means that it may not be possible to study individual variations using visualization methods on deep learning applied to imaging data. This problem might be resolved by using more samples and with a better initialization, given for example by an autoencoder pretraining. Moreover, we found that the regions identified by the networks were very small (restricted to the hippocampus, amygdala and part of the temporal lobe) and that most of the image is not exploited by the CNN to find the diagnosis. This confirms the findings of [17] in which an equal performance was found by learning AD vs CN classification with a CNN on the whole MRI or the hippocampus only. This focus of the CNN on the hippocampi only may be partly due to the dataset: ADNI is a research cohort from which patients with multiple phenotypes are excluded, leading to a very homogeneous cohort in which the main symptom of the patients is memory loss. It does not fully represent the diversity of AD phenotype, and they may be biased towards hippocampus atrophy.

The visualization method we used has several limitations. First, the quality check of the data is crucial otherwise the training of the group level mask is not stable. Second, our method is only meaningful for quantitative data: for T1-MRI it would not have been sensible to increase the value of the voxels as it would have deformed the image in a non-meaningful way. This is an issue as the advantage of deep learning is precisely to be able to adapt to the rawest data possible. Finally, though we explored the effect of four hyperparameters of the visualization method ($\beta_1$, $\beta_2$, $\lambda_1$, $\lambda_2$) we did not conduct an exhaustive study on the impact of the learning rate and the number of epochs performed (correlated to the patience, the tolerance and the maximum number of epochs). As we have seen when comparing masks trained on ADNI and AIBL, these parameters impact the amount of regularization of the masks.

Future work will consist of developing a mask for non-quantitative images by creating another version of the input that is sensible. For example, for AD classification with T1-MRI, we would need the healthy version of an AD subject to mask the image with the healthy version. This kind of transformation from AD to normality has been explored in [18]. Moreover, we would like to study whether the visualization method is able to differentiate between overfitting, random and correctly trained networks. Finally, we will explore how masks can be used to enhance the training performance, for example by following an adversarial strategy to force the network to base its decision on more various regions.

# ACKNOWLEDGEMENTS

The research leading to these results has received funding from the French government under management of Agence Nationale de la Recherche as part of the "Investissements d'avenir" program, reference ANR-19-P3IA-0001 (PRAIRIE 3IA Institute) and reference ANR-10-IAIHU-06 (Agence Nationale de la Recherche-10-IA Institut Hospitalo-Universitaire-6).

All experiments were performed on the cluster of the Brain and Spine Institute in Paris, which is equipped with 4 NVIDIA P100 GPU cards (64 GB shared memory) and 24 CPUs (120 GB shared memory).

# REFERENCES

[1]     G. Litjens *et al.*, 'A survey on deep learning in medical image analysis', *Med. Image Anal.*, vol. 42, pp. 60–88, Dec. 2017.

[2]     K. Simonyan, A. Vedaldi, and A. Zisserman, 'Deep Inside Convolutional Networks: Visualising Image Classification Models and Saliency Maps', *ArXiv13126034 Cs*, Dec. 2013.

[3]     J. T. Springenberg, A. Dosovitskiy, T. Brox, and M. Riedmiller, 'Striving for Simplicity: The All Convolutional Net', *ArXiv14126806 Cs*, Dec. 2014.

[4]     J. Zhang, S. A. Bargal, Z. Lin, J. Brandt, X. Shen, and S. Sclaroff, 'Top-Down Neural Attention by Excitation Backprop', *Int. J. Comput. Vis.*, vol. 126, no. 10, pp. 1084–1102, Oct. 2018.

[5]     R. R. Selvaraju, M. Cogswell, A. Das, R. Vedantam, D. Parikh, and D. Batra, 'Grad-CAM: Visual Explanations from Deep Networks via Gradient-Based Localization', in *2017 IEEE International Conference on Computer Vision (ICCV)*, 2017, pp. 618–626.

[6]     M. D. Zeiler and R. Fergus, 'Visualizing and Understanding Convolutional Networks', in *Computer Vision – ECCV 2014*, 2014, pp. 818–833.

[7]     M. T. Ribeiro, S. Singh, and C. Guestrin, '"Why Should I Trust You?": Explaining the Predictions of Any Classifier', in *Proceedings of the 22nd ACM SIGKDD International Conference on Knowledge Discovery and Data*

*Mining - KDD '16*, San Francisco, California, USA, 2016, pp. 1135–1144.

[8] R. C. Fong and A. Vedaldi, 'Interpretable Explanations of Black Boxes by Meaningful Perturbation', in *2017 IEEE International Conference on Computer Vision (ICCV)*, 2017, pp. 3449–3457.

[9] A. Routier *et al.*, 'Clinica: an open source software platform for reproducible clinical neuroscience studies', Jun. 2018.

[10] J. Samper-González *et al.*, 'Reproducible evaluation of classification methods in Alzheimer's disease: Framework and application to MRI and PET data', *NeuroImage*, vol. 183, pp. 504–521, Dec. 2018.

[11] J. Ashburner and K. J. Friston, 'Unified segmentation', *NeuroImage*, vol. 26, no. 3, pp. 839–851, Jul. 2005.

[12] J. Bergstra and Y. Bengio, 'Random Search for Hyper-Parameter Optimization', *J. Mach. Learn. Res.*, vol. 13, no. Feb, pp. 281–305, 2012.

[13] K. He, X. Zhang, S. Ren, and J. Sun, 'Delving Deep into Rectifiers: Surpassing Human-Level Performance on ImageNet Classification', in *2015 IEEE International Conference on Computer Vision (ICCV)*, Santiago, Chile, 2015, pp. 1026–1034.

[14] E. T. Rolls, M. Joliot, and N. Tzourio-Mazoyer, 'Implementation of a new parcellation of the orbitofrontal cortex in the automated anatomical labeling atlas', *NeuroImage*, vol. 122, pp. 1–5, Nov. 2015.

[15] J. Adebayo, J. Gilmer, M. Muelly, I. Goodfellow, M. Hardt, and B. Kim, 'Sanity checks for saliency maps', in *Advances in Neural Information Processing Systems*, 2018, pp. 9505–9515.

[16] J. L. Whitwell *et al.*, '3D maps from multiple MRI illustrate changing atrophy patterns as subjects progress from mild cognitive impairment to Alzheimer's disease', *Brain*, vol. 130, no. 7, pp. 1777–1786, Jul. 2007.

[17] J. Wen *et al.*, 'Convolutional Neural Networks for Classification of Alzheimer's Disease: Overview and Reproducible Evaluation', *ArXiv190407773 Cs Eess Stat*, Apr. 2019.

[18] C. Bowles, R. Gunn, A. Hammers, and D. Rueckert, 'Modelling the progression of Alzheimer's disease in MRI using generative adversarial networks', in *Medical Imaging 2018: Image Processing*, Houston, United States, 2018, p. 55.